# Validation approaches for satellite-based PM$_{2.5}$ estimation:

# Assessment and a new approach

Tongwen Li [a], Huanfeng Shen [a,b,c*], Qiangqiang Yuan [d,b], Liangpei Zhang [e,b]

[a] School of Resource and Environmental Sciences, Wuhan University, Wuhan, 430079, China.

[b] Collaborative Innovation Center of Geospatial Technology, Wuhan, 430079, China.

[c] The Key Laboratory of Geographic Information System, Ministry of Education, Wuhan University, Wuhan, 430079, China.

[d] School of Geodesy and Geomatics, Wuhan University, Wuhan, 430079, China.

[e] The State Key Laboratory of Information Engineering in Surveying, Mapping and Remote Sensing, Wuhan University, Wuhan, 430079, China.

\* Corresponding author. E-mail address: shenhf@whu.edu.cn

**Abstract**

Satellite-derived aerosol optical depth (AOD) has been increasingly employed for the estimation of ground-level PM$_{2.5}$, which is often achieved by modeling the relationship between AOD and PM$_{2.5}$. To evaluate the accuracy of PM$_{2.5}$ estimation, the cross-validation (CV) technique has been widely used. There have been several CV-based validation approaches applied for the AOD-PM$_{2.5}$ models. However, the applicable conditions of these validation approaches still remain unclear. Additionally, is there space to develop better validation approaches for the AOD-PM$_{2.5}$ models? The contributions of this study can be summarized as two aspects. Firstly, we comprehensively analyze and assess the existing validation approaches, and give the suggestions for applicable conditions of them. Then, the existing validation approaches do not take the distance to monitoring station into consideration. A new validation approach considering the distance to monitoring station is proposed in this study. Among the existing validation approaches, the sample-based CV is used to reflect the overall prediction ability; the site-based CV and the region-based CV have the potentials to evaluate spatial prediction performance; the time-based CV and the historical validation are capable of evaluating temporal prediction accuracy. In addition, the validation results indicate that the proposed validation approach has shown great potentials to better evaluate the accuracy of $PM_{2.5}$ estimation. This study provides application implications and new perspectives for the validation of AOD-$PM_{2.5}$ models.

## 1. Introduction

With the rapid development of economy, air pollution has aroused worldwide concerns in recent years. As reported by a study of World Health Organization (WHO), the public health is heavily influenced by air pollution during the 21th century (WHO, 2006). Thereinto, fine particulate matter ($PM_{2.5}$, particulate matters with aerodynamic diameters of less than 2.5 $\mu m$) pollution has gradually become one of the main air pollutants (Engel-Cox et al., 2013; Peng et al., 2016). To monitor the $PM_{2.5}$ pollution, the ground stations are considered to be the most reliable with high accuracy of $PM_{2.5}$ measurements. However, due to the high cost of monitoring stations, the $PM_{2.5}$ station network is usually sparsely distributed in space.

Satellite remote sensing has been widely employed to extend the $PM_{2.5}$ monitoring beyond the ground stations, owing to its large temporal and spatial coverage. To estimate $PM_{2.5}$ from satellite remote sensing, the most widely used strategy is establish a statistical relationship model between satellite-derived aerosol optical depth (AOD) and ground-level $PM_{2.5}$. Hence, there have been many AOD-$PM_{2.5}$ models developed for the estimation of $PM_{2.5}$ from satellite observations. These AOD-PM$_{2.5}$ models mainly include the early statistical models, such as, multiple linear regression (Gupta and Christopher, 2009b), semi-empirical model (Liu et al., 2005; Tian and Chen, 2010), and so on, and more advanced statistical models, for instance, geographically weighted regression (Hu et al., 2013), linear mixed effects model (Lee et al., 2011), neural networks (Gupta and Christopher, 2009a; Li et al., 2017b), and etc. However, how to validate the AOD-PM$_{2.5}$ models still remains a challenge.

To evaluate the estimation accuracy of AOD-PM$_{2.5}$ models, the station PM$_{2.5}$ measurements are commonly adopted. Hence, a cross-validation (CV) technique (Rodriguez et al., 2010) is often applied, which in fact leaves some station PM$_{2.5}$ observations for model validation. Based on the CV technique, several validation approaches have been employed for the model validation, including the sample-based CV (He and Huang, 2018; Li et al., 2017b), the site-based CV (Lee et al., 2011; Xie et al., 2015), the region-based CV (Li et al., 2017a), the time-based CV (Ma et al., 2016b; Zhang et al., 2018), and so on. In addition, some studies focused on the historical prediction of PM$_{2.5}$ concentrations. The historical validation (Huang et al., 2018; Ma et al., 2016a), which does not use a CV strategy, has also been exploited. Some studies focusing on satellite-based PM$_{2.5}$ estimation adopted only one of the above-mentioned validation approaches, while some studies simultaneously used several validation approaches.

However, the existing validation approaches often make the readers confused, because a same AOD-PM$_{2.5}$ model may report different validation results using different validation approaches. The applicable conditions of these validation approaches still remain unclear. Researchers often did not choose an appropriate validation approach in accordance with their study goals. Therefore, it is of great importance to clarify the applicable conditions of the existing validation approaches. On the other hand, the monitoring stations are often established in the centers of cities. The validation stations are usually very close to the modeling stations. As a result, previous validation approaches may just evaluate the estimation accuracy of locations close to monitoring stations, but have no capacities to reflect the estimation accuracy of locations father to monitoring stations. They do not take the distance to monitoring stations into consideration, which will bring some biases for the evaluation of AOD-PM$_{2.5}$ models. Is there space to develop a better validation approach for AOD-PM$_{2.5}$ models?

In this paper, one of the main objectives is to comprehensively analyze and assess the existing validation approaches. On the basis, this study will give some suggestions for the applicable conditions of the existing validation approaches. Secondly, a new validation approach considering the distance to monitoring stations will be developed. The rest of this paper is organized as follows. Section 2 is going to fully analyze the existing validation approaches. In section 3, we will propose a new validation approach, and it will be verified by a case study. The discussions and conclusions will be summarized in section 4.

**2. Previous validation approaches**

Using ground station measurements to validate the estimates from satellite remote sensing is a common strategy. Hence, the cross-validation technique, which actually leaves some station observations for model validation, has been widely adopted to validate the AOD-PM$_{2.5}$ models. For the $k$-fold cross-validation (Rodriguez et al., 2010), the samples (stations, regions, or times) are divided into $k$ folds randomly and averagely. Then, $k-1$ folds of them are used for model fitting and the remaining one used for model validation. Finally, the above process will be repeated $k$ times, so as to validate the model performance on each fold. Thereinto, $k$ is often set as 10 or the number of samples (stations, regions, or times), which indicates the 10-fold CV or the leave-one CV, respectively. In the field of AOD-PM$_{2.5}$ studies, for cross-validation, the input data can be data samples, monitoring sites, stations in one region, or stations in one time, which are named as sample-, site-, region-, or time-based CV, respectively. In addition, the historic validation, which does not belong to a cross-validation technique, has also been used for validating the AOD-PM$_{2.5}$ models. The schematics of these validation approaches are presented in Figure 1, and the details of them are as follows.

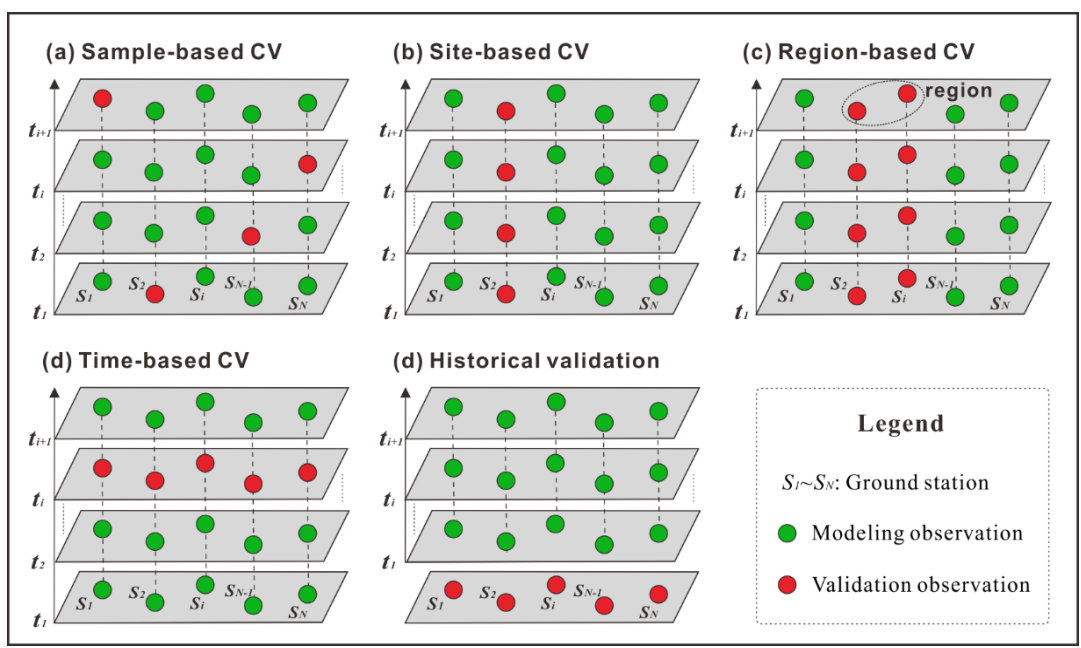

**Figure 1.** Schematics of various validation approaches. (a)-(d) only present one round of validation.

*2.1. The sample-based CV*

As presented in Figure 1(a), we mix up the locations and times of the AOD-PM$_{2.5}$ matchup samples, and then randomly select some samples for cross-validation. Please note that, for a certain monitoring station, the samples from this station on some times are used for modeling fitting, and samples on the other times for model validation. That means the modeling dataset and validation dataset may contain the same monitoring stations. In general, the sample-based

CV approach is often employed to reflect the overall prediction ability of AOD-PM$_{2.5}$ models. However, this validation approach has some limitations, that is, a same monitoring station may be simultaneously used for the model fitting and the model validation, i.e., some times of observations for model fitting, and the other times of observations for model validation. This brings some biases to evaluate the model prediction ability for satellite-based mapping of ground-level PM$_{2.5}$, because the locations with PM$_{2.5}$ values to be estimated have no ground stations in the real situations.

*2.2. The site-based CV*

Unlike the sample-based CV, the monitoring stations are randomly selected for model validation in the site-based CV approach, which is shown in Figure 1(b). For the site-based CV, the validation stations are never included in the model fitting, that is, the validation station samples on all times are merely for model validation. Hence, the site-based CV has the potentials to evaluate the accuracy of PM$_{2.5}$ spatial prediction. Here, it should be noted that the site-based CV often refers to the grid cell-based CV, because the PM$_{2.5}$ values from multiple monitoring stations in a grid cell are often averaged in the model development (Li et al., 2017a; Shen et al., 2017). However, as is well known, the monitoring stations are usually established in the center of cities, and the vicinities of validation stations often exist modeling stations. Thus, the site-based CV has some limitations in the evaluation of model performance, which implies the site-based CV is not able to reflect the prediction accuracy of PM$_{2.5}$ on the locations with farther distance to monitoring stations. Although the site-based CV is a spatial-out validation approach, but it tends to only reflect the prediction accuracy of locations close to monitoring stations.

*2.3. The region-based CV*

To some degrees, the region-based CV has the potentials to avoid the limitations of the site-based CV. As observed from Figure 1(c), some certain regions, e.g., a province (Li et al., 2017a), are chosen for model validation. So the region-based CV approach also adopts a spatial-out strategy like the site-based CV. The stations in the validation region are all used for model validation, and it may be capable of evaluating the spatial prediction accuracy of $PM_{2.5}$ on the locations with farther distances to the monitoring stations. However, what is the optical extent for the validation regions? In consider of the uneven distribution of monitoring stations, it still meets some challenges to select a reasonable extent for model validation.

*2.4. The time-based CV*

Then follows by the time-based CV approach, which is illustrated in Figure 1(d). Unlike the above validation approaches, which pay more attentions on the evaluation of spatial prediction, the time-based CV is often used to evaluate the accuracy of temporal prediction. Under some situations, the satellite-derived AOD data are available, whereas the station $PM_{2.5}$ observations are absent. The AOD-$PM_{2.5}$ models are established using samples on the times with both AOD and $PM_{2.5}$ data, how well do the models perform on those times without AOD-$PM_{2.5}$ matchups? Hence, we randomly choose some times of observations for model validation, and the remaining times of observations are used for model fitting. The time-based CV is expected to evaluate the prediction accuracy for those times without AOD-$PM_{2.5}$ matchups. However, in the real situations, the times with satellite AOD but without $PM_{2.5}$ measurements are relatively rare. Therefore, due to the a few applicable conditions, the use of the time-based CV is limited for the assessment of the AOD-$PM_{2.5}$ models.

*2.5. The historical validation*

With a large temporal coverage of satellite observations, the AOD-PM$_{2.5}$ models have shown the capacities to predict historical PM$_{2.5}$ concentrations. Hence, the historical validation approach is developed to evaluate the accuracy of historical prediction. As presented in Figure 1(e), long time historical PM$_{2.5}$ data are collected for the validation of AOD-PM$_{2.5}$ models. Here, the main differences between the historical validation and the time-based CV are that, the historical validation uses long time historical PM$_{2.5}$ data, while the time-based CV randomly chooses some times of PM$_{2.5}$ observations in the study period for cross-validation. It should be noted that some studies used future PM$_{2.5}$ data for historical validation, for example, the model was established using samples from 2013, and was validated using PM$_{2.5}$ data in the first half of 2014. Then the model was employed to predict historical PM$_{2.5}$ values on 2004-2012 (Ma et al., 2016a). The limitations for the historical validation approach are that, it is often hard to collect sufficient PM$_{2.5}$ data for validation.

**Table 1**. Summary of various validation approaches

| Approach | Applicable conditions | Limitations |
| --- | --- | --- |
| The sample-based CV | The overall prediction accuracy | Modeling dataset and validation dataset may contain the same stations |
| The site-based CV | The spatial prediction accuracy | Tends to only reflect the prediction accuracy of locations close to stations |
| The region-based CV | The spatial (region) prediction accuracy | The selection of the optical extent for the validation region is uncertain |
| The time-based CV | The temporal prediction accuracy | The real situations with satellite AOD but without PM$_{2.5}$ observations are rare |
| Historical validation | The temporal (historical) prediction accuracy | May be hard to collect sufficient historical PM$_{2.5}$ data |

To sum up, it can be concluded from the above analyses that, the sample-based CV has the ability to evaluate the overall accuracy of PM$_{2.5}$ estimation; the site-based CV and the region-based CV are expected to evaluate the spatial prediction accuracy; the time-based CV and the historical validation are used for the evaluation of temporal prediction accuracy. On the basis, we give some suggestions for the applicable conditions and limitations for these validation approaches, which can be seen in Table 1.

**3. The proposed validation approach**

*3.1. The development of the proposed validation approach*

As explained in the site-based CV, the monitoring stations are unevenly established in space, for example, mainly in the centers of cities. Hence, the stations are often distributed sparsely in a large range and relatively densely in a small range. Thus, the monitoring station is close to its neighbors, as a result, the site-based CV only reflects the prediction accuracy on the locations near the stations. However, the performance on locations farther to the stations still remains unclear. On the other side, due to the uneven distribution of monitoring stations, the existing validation approaches may have the risks to misjudge the superiorities of the AOD-PM$_{2.5}$ methods. For instance, the spatial interpolation may achieve a better PM$_{2.5}$ estimation result, due to the closeness of validation stations and modeling stations. Therefore, it is of great significance to develop a validation approach considering the distance to stations for a better evaluation of the AOD-PM$_{2.5}$ models.

Supposing that $m$ validation stations are collected, which forms a validation collection $\mathbf{S}_{val} = \{S_{v,1}, S_{v,2}, ..., S_{v,m}\}$, and accordingly, $\mathbf{S}_{fit} = \{S_{f,1}, S_{f,2}, ..., S_{f,n}\}$ denotes the modeling collection with $n$ monitoring stations, where $S$ means the monitoring station, the total number of monitoring stations is $N = m + n$. To evaluate the performance on space locations, we manually set a distance of $d$ ($km$). For a given distance $d$, the modeling stations with a distance to any validation station of less than $d$ are excluded from modeling collection $\mathbf{S}_{fit}$ (see step 2 in Figure 2). Thus, the distances of validation stations to modeling stations are all greater than $d$. Then, the AOD-PM$_{2.5}$ estimation model considering the distance to stations can be depicted as Equation (1).

$$PM_{d\text{-}fit} = f_{(d)}(X_{d\text{-}fit}) \tag{1}$$

where $X_{d\text{-}fit}$ is input variables (e.g. AOD, meteorological data, etc.) of the modeling dataset for the distance $d$, and $f_{(d)}$ refers to the estimation function considering the distance $d$. Subsequently, the AOD-PM$_{2.5}$ model can be validated using the validation collection, which is represented as Equation (2).

$$PM_{d\text{-}val} = f_{(d)}(X_{d\text{-}val}) \tag{2}$$

where $X_{d\text{-}val}$ is input variables of the validation dataset for the distance $d$. Through the calculation of equation (2), the estimated PM$_{2.5}$ values can be obtained. Comparing the estimated PM$_{2.5}$ with station PM$_{2.5}$ measurements, the model performance will be evaluated.

Figure 2 presents the workflow of the proposed validation approach, which mainly consists of three steps.

*Step 1*: Based on a 10-fold CV, the monitoring stations are divided into 10 folds randomly and averagely, for each validation fold, we can obtain the validation collection of stations ($\mathbf{S}_{val}$), and the remaining stations are used as modeling collection ($\mathbf{S}_{fit}$).

*Step 2*: We manually set a distance $d$, and the modeling collection will be updated according to the distance from validation station to modeling station. Using the updated modeling dataset, the AOD-PM$_{2.5}$ model can be trained.

*Step 3*: Through the established AOD-PM$_{2.5}$ model, the estimated PM$_{2.5}$ values will be calculated and compared against the station PM$_{2.5}$ measurements in the validation dataset. Thus, the model performance can be evaluated.

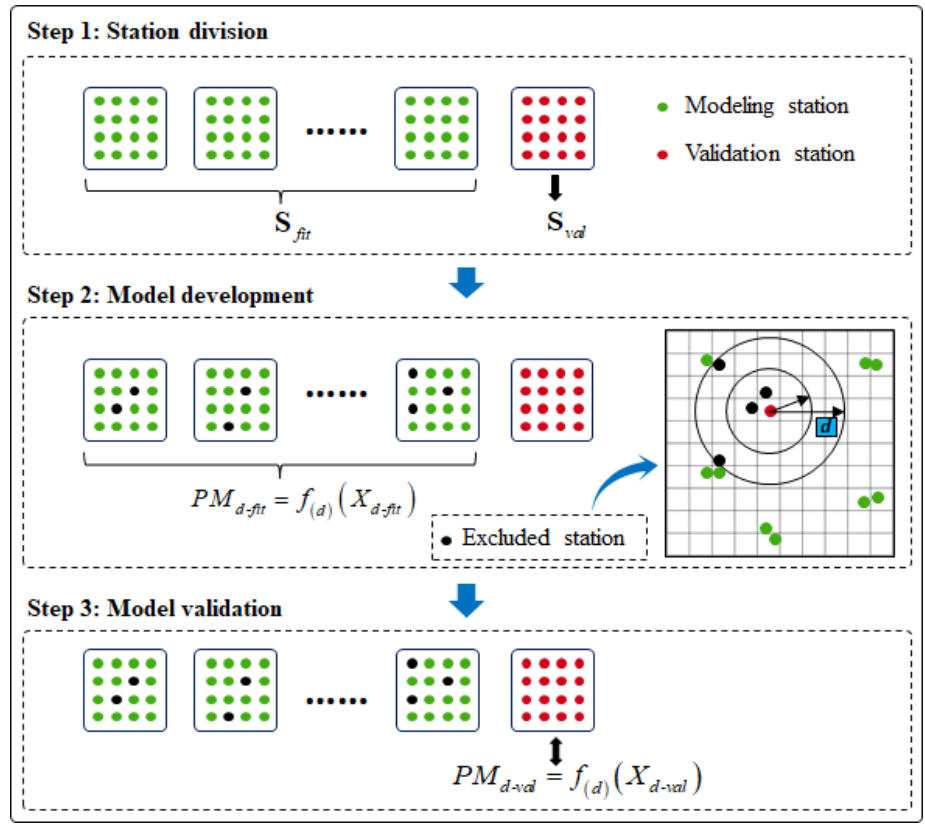

**Figure 2**. Schematic and flowchart of the proposed validation approach.

*3.2. A case study using the proposed validation approach*

*3.2.1. Study region and data*

To verify the proposed validation approach, a case study is conducted on the region of China. The study region is shown in Figure 3, and a total of ~1500 monitoring stations have been collected for the model development and validation. The study period is from January 1, 2015 to December 31, 2015. The annual mean PM$_{2.5}$ values on each station were calculated, and presented in Figure 3.

The data used in this study include four main parts. Briefly, they are described as follows. 1) Ground-level PM$_{2.5}$. We collected hourly PM$_{2.5}$ measurements from the China National

Environmental Monitoring Center (CNEMC) website (http://106.37.208.233:20035/). The hourly PM$_{2.5}$ data were averaged to daily mean PM$_{2.5}$ in this study. 2) MODIS AOD. Both Terra and Aqua MODIS AOD products were downloaded from the Level 1 and Atmosphere Archive and Distribution System (LAADS, https://ladsweb.modaps.eosdis.nasa.gov/). We exploited Collection 6 AOD products, which are retrieved by combining dark target and deep blue algorithms (Levy et al., 2013). They have a spatial resolution of 10 km. The average of the Terra and Aqua AOD products was employed to estimate daily PM$_{2.5}$. 3) Meteorological parameters. We extracted wind speed at 10 m above ground (WS, m/s), air temperature at a 2 m height (TMP, K), relative humidity (RH, %), surface pressure (PS, kPa), and planetary boundary layer height (PBL, m) from MERRA-2 meteorological reanalysis data, which were downloaded from the NASA website (http://gmao.gsfc.nasa.gov/GMAO_products/). 4) MODIS normalized difference vegetation index (NDVI). MODIS NDVI products (MOD13) were also achieved from the LAADS website. The details of data used can refer to our previous study (Li et al., 2017a).

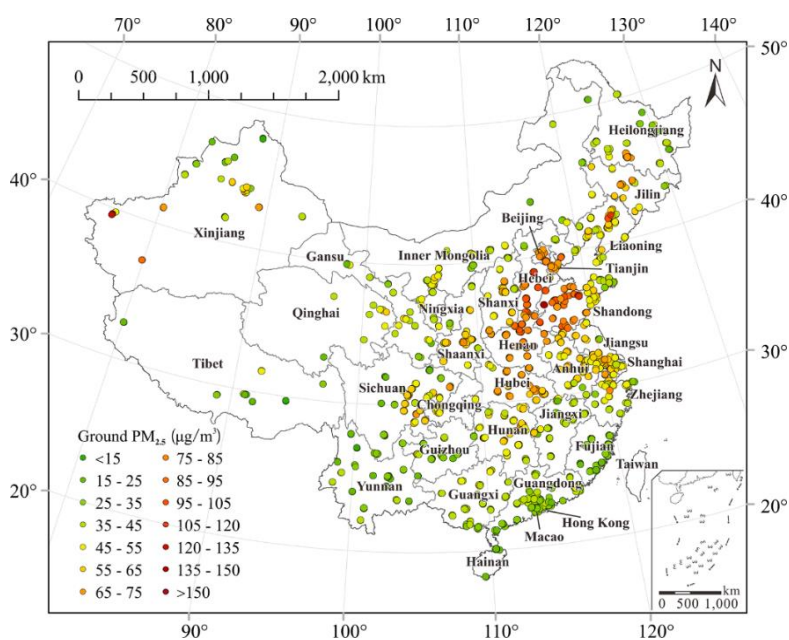

**Figure 3**. Study region and the spatial distribution of PM$_{2.5}$ stations.

*3.2.2. The AOD-PM$_{2.5}$ model*

The proposed validation approach is used to evaluate the performance of the Geoi-DBN model, which was developed in our previous study (Li et al., 2017a). In brief, the Geoi-DBN model represents a relationship between PM$_{2.5}$, satellite AOD, meteorological factors, satellite NDVI, and geographical correlation of PM$_{2.5}$. The procedure of the Geoi-DBN for the satellite-based estimation of ground PM$_{2.5}$ consists of three steps.

*Step 1:* The input variables (satellite AOD, meteorological factors, satellite NDVI, and geographical correlation of PM$_{2.5}$) are input into the Geoi-DBN model. This model is pre-trained without supervision to initialize itself. That is, the station PM$_{2.5}$ measurements are not used in this step, the initial model coefficients are trained from input data.

*Step 2:* The model-estimated PM$_{2.5}$ can be obtained. Subsequently, we calculate mean square error between estimated PM$_{2.5}$ and ground observed PM$_{2.5}$. The error is sent back to deep learning model to fine-tune the model coefficients using back-propagation (BP) algorithm (Rumelhart et al., 1986). This process will be repeated until the Geoi-DBN model achieves a satisfactory performance.

*Step 3:* The Geoi-DBN model will be evaluated and utilized to predict the spatial PM$_{2.5}$ values where there is no monitoring stations. Thus, the spatial distribution of PM$_{2.5}$ concentrations can be obtained.

*3.2.3. Results and analyses*

To evaluate the Geoi-DBN model using the proposed validation approach, we set the distance $d$ with the bounds of 0-110 km and the step of 10 km. The statistical indexes used include determination coefficient (R$^2$, unitless), root-mean-square error (RMSE, $\mu g/m^3$), mean predictive error (MPE, $\mu g / m^3$), and relative prediction error (RPE, %). The validation results are shown in Table 2.

**Table 2**. The validation results of Geoi-DBN model using the proposed validation approach

| $d$ (km) | $R^2$ (unitless) | RMSE ($\mu g / m^3$) | MPE ($\mu g / m^3$) | RPE (%) |
|---|---|---|---|---|
| 0   | 0.84 | 15.39 | 9.87  | 28.06 |
| 10  | 0.80 | 17.11 | 11.16 | 31.18 |
| 20  | 0.72 | 20.33 | 13.45 | 37.05 |
| 30  | 0.66 | 22.33 | 14.56 | 40.70 |
| 40  | 0.68 | 21.74 | 14.56 | 39.63 |
| 50  | 0.65 | 22.59 | 14.98 | 41.17 |
| 60  | 0.65 | 22.77 | 15.14 | 41.50 |
| 70  | 0.63 | 23.22 | 15.41 | 42.33 |
| 80  | 0.63 | 23.29 | 15.64 | 42.46 |
| 90  | 0.61 | 24.16 | 16.03 | 44.03 |
| 100 | 0.59 | 24.53 | 16.40 | 44.72 |
| 110 | 0.58 | 24.90 | 16.69 | 45.40 |

As presented in Table 2, when the distance is set as 0 km, the proposed validation approach is in fact the site-based CV. The $R^2$ and RMSE values are 0.84 and 15.39 $\mu g / m^3$, respectively. Here, the performance differs from that of our previous study, the reason for this is that the Geoi-DBN model is slightly improved by calculating temporal correlation term using spatial correlation term rather than the historical PM$_{2.5}$ from the same station. As the distance increases, the validation stations get farther to the modeling stations. Overall, the model performance is gradually decreasing, with $R^2$ values decreasing from 0.84 for 0 km to 0.58 for 110 km.

To characterize the decreasing treads of model performance, we fit a linear regression between $R^2$ and distance, which is exhibited in Figure 4. The relationship between $R^2$ and distance is well fitted using a two-stage linear regression, with goodness-of-fit $R^2$ values of 0.99 and 0.95 respectively. The slopes of the two stages are -0.0062 and -0.0015, respectively, indicating that the performance of Geoi-DBN model decreases relatively sharply in the first stage (0-30 km) and more tardy in the second stage. Through the validation using the proposed approach, how the Geoi-DBN model performs with farther distance to stations is reasonably evaluated. Therefore, the validation results show that the proposed validation approach has shown the superiority to fully evaluate the AOD-PM$_{2.5}$ model.

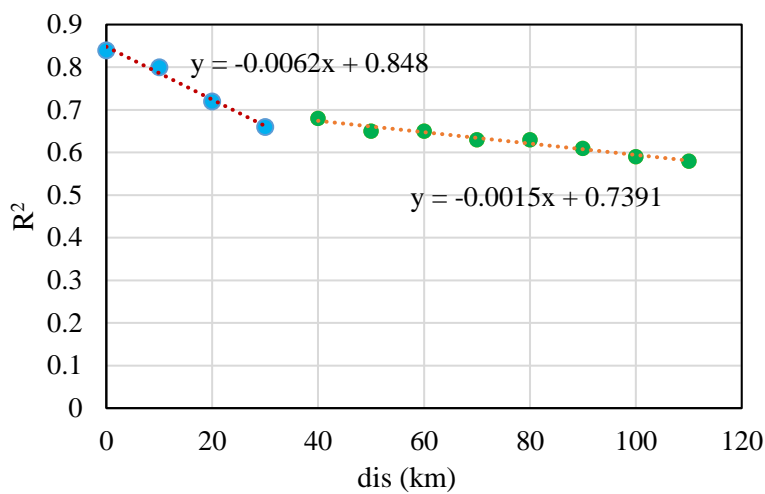

Figure 4. The trends of model performance with increasing distance.

## 4. Discussions and conclusions

To sum up, this work has two aspects of contributions to the field AOD-PM$_{2.5}$ studies. Firstly, there have been many validation approaches for the AOD-PM$_{2.5}$ models, however, these validation approaches often make the readers confused. Because a same AOD-PM$_{2.5}$ model often shows greatly different validation results, while the applicable conditions of these validation approaches still remain unclear. Hence, one important contribution of this study is that we fully analyze and assess the existing validation approaches, and give some suggestions for the applicable conditions of them. The researchers can select the appropriate validation approaches according to the goal of their studies. In addition, the existing validation approaches do not consider the distances to monitoring stations, which may bring some biases for the evaluation of AOD-PM$_{2.5}$ models. A novel validation approach, which takes the distances to stations, is proposed in this study. The results show that the proposed validation approach has the capacities to fully evaluate the AOD-PM$_{2.5}$ models. In a word, this study provides application implications and new perspectives for the validation of AOD-PM$_{2.5}$ models.

**Acknowledgements**

This study was funded by the National Key R&D Program of China (No. 2016YFC0200900). We are grateful to the China National Environmental Monitoring Center (CNEMC, http://106.37.208.233:20035/), the Level 1 and Atmosphere Archive and Distribution System (LAADS, https://ladsweb.modaps.eosdis.nasa.gov/), and the Goddard Space Flight Center Distributed Active Archive Center (GSFC DAAC, http://gmao.gsfc.nasa.gov/GMAO_products/) for providing ground PM$_{2.5}$ measurements, satellite data, and meteorological data, respectively.